\documentclass[fleqn,usenatbib,twocolumn]{aastex631}

\usepackage{newtxtext,newtxmath}

\usepackage[T1]{fontenc}

\DeclareRobustCommand{\VAN}[3]{#2}
\let\VANthebibliography\thebibliography
\def\thebibliography{\DeclareRobustCommand{\VAN}[3]{##3}\VANthebibliography}

\usepackage{graphicx}	
\usepackage{amsmath}	

\usepackage{xcolor}

\usepackage[normalem]{ulem}

\newcommand{\alphalo}{\alpha_{\ast,\text{lo}}}
\newcommand{\alphahi}{\alpha_{\ast,\text{hi}}}
\newcommand{\xHI}{x_{\text{H } \textsc{i}}}

\begin{document}

\title{The relative constraining power of the high-$z$ 21-cm dipole and monopole signals}

\shortauthors{Mirocha et al.}
\shorttitle{21-cm dipole vs. monopole}

\author[0000-0002-8802-5581]{Jordan Mirocha}
\affiliation{Jet Propulsion Laboratory, California Institute of Technology, 4800 Oak Grove Drive, Pasadena, CA 91109, USA}
\affiliation{California Institute of Technology,  1200 E. California Boulevard, Pasadena, CA 91125, USA}

\author[0000-0001-6156-4449]{Chris Anderson}
\affiliation{Jet Propulsion Laboratory, California Institute of Technology, 4800 Oak Grove Drive, Pasadena, CA 91109, USA}
\affiliation{California Institute of Technology, 1200 E. California Boulevard, Pasadena, CA 91125, USA}

\author[0000-0001-5929-4187]{Tzu-Ching Chang}
\affiliation{Jet Propulsion Laboratory, California Institute of Technology, 4800 Oak Grove Drive, Pasadena, CA 91109, USA}
\affiliation{California Institute of Technology, 1200 E. California Boulevard, Pasadena, CA 91125, USA}

\author[0000-0001-7432-2932]{Olivier Dor\'{e}}
\affiliation{Jet Propulsion Laboratory, California Institute of Technology, 4800 Oak Grove Drive, Pasadena, CA 91109, USA}
\affiliation{California Institute of Technology, 1200 E. California Boulevard, Pasadena, CA 91125, USA}

\author[0000-0002-3950-9598]{Adam Lidz}
\affiliation{Center for Particle Cosmology, Department of Physics and Astronomy, University of Pennsylvania, Philadelphia, PA 19104, USA}

\begin{abstract}
The 21-cm background is a promising probe of early star formation and black hole activity. While a slew of experiments on the ground seek to detect the 21-cm monopole and spatial fluctuations on large $\sim 10$ arcminute scales, little work has been done on the prospects for detecting the 21-cm dipole signal or its utility as a probe of early galaxies. Though an intrinsically weak signal relative to the monopole, its direction is known well from the cosmic microwave background and wide-field surveys, plus as a relative measurement the dipole could help relax instrumental requirements. In order to understand the constraining power of the dipole, in this work we perform parameter inference on mock datasets that include the dipole, monopole, or both signals. We find that while the monopole does provide the best constraints for a given integration time, constraints from a dipole measurement are competitive, and can in principle constrain the cosmic star formation rate density and efficiency of X-ray photon production in early $z \sim 15$ galaxies to better than a factor of $\sim 2$. This result holds for most of the available prior volume, which is set by constraints on galaxy luminosity functions, the reionization history, and upper limits from 21-cm power spectrum experiments. We also find that predictions for the monopole from a dipole measurement are robust to different choices of signal model. As a result, the 21-cm dipole signal is a valuable target for future observations and offers a robust cross-check on monopole measurements.
\end{abstract}



\section{Introduction} \label{sec:intro}
Observations of the cosmic 21-cm background from redshifts $z \gtrsim 6$ are a powerful probe of the Epoch of Reionization (EoR) and cosmic dawn, when the first stars and galaxies began to transform their environments through ionization and X-ray heating \citep{Madau1997,FurlanettoOhBriggs2006,Morales2010,Pritchard2012}. A fleet of arrays on the ground are currently seeking a detection of the 21-cm power spectrum during the EoR \citep[LOFAR, MWA, HERA, GMRT, LWA;][]{vanHaarlem2013,Tingay2013,DeBoer2017,Paciga2013,Eastwood2019}, while in parallel, a suite of more modest single-element receivers \citep{Bowman2010,Singh2017,REACH,Philip2019,Monsalve2023} are pursuing a detection of the sky-averaged ``global'' 21-cm signal \citep{Shaver1999}, which traces the mean properties of the IGM at early times rather than spatial fluctuations. Both measurements encode a wealth of information on early star and black hole formation \citep[e.g.,][]{Mesinger2013,Fialkov2014HMXBs}, as well as the intergalactic medium \citep[IGM; e.g.,][]{Cohen2017,Mirocha2022,Ghara2024}, and are thus a powerful complement to high-$z$ galaxy surveys \citep[e.g.,][]{Mirocha2017,Park2019,Hutter2021,Ma2023}. In the long term, higher order statistics and 21-cm maps promise to deliver even more detailed information about the cosmic dawn \citep[e.g.,][]{Lidz2007,Watkinson2019,LaPlante2019,Greig2022,Gillet2019,Hassan2020,Zhao2022}.

In the last $\sim 10$ years there has been tremendous progress in efforts to detect the 21-cm background. The EDGES collaboration reported the detection of a feature in the sky-averaged spectrum \citep{Bowman2018}, potentially consistent with expectations for the global 21-cm signal, but strong enough to drive a considerable flurry of activity in the modeling community. Meanwhile, upper limits from power spectrum measurements have continued to improve in the last few years \citep{Mertens2020,MWA2020Obs,HERA2022Obs}, and have recently breached the parameter space of ``normal'' models, i.e., those that do not invoke exotic mechanisms in order to amplify fluctuations beyond the theoretical maximum set by a $\Lambda$CDM cosmology. Arrays operating at higher frequencies, and so targeting neutral hydrogen in the post-reionization Universe, have reported auto-correlation and cross-correlation detections \citep{Chang2010,CHIME2023,MEERKAT2023}, demonstrating both a maturity in the calibration and analyses of low-frequency radio interferometers \citep[see review by][]{LiuShaw2020}, and detection proof-of-principle, albeit in a frequency regime where foregrounds are weaker than those relevant to EoR studies.

Despite rapid experimental and theoretical progress focused on the 21-cm monopole and power spectrum, relatively little work has been done on the 21-cm dipole signal. As first discussed in \citet{Slosar2017}, the 21-cm signal should show a kinematic dipole spatial variation owing to our motion with respect to the frame of the emitting or absorbing hydrogen gas. Though $\sim 10^2$ times weaker than the monopole, \citet{Slosar2017} pointed out that the dipole may be an appealing target for near-future experiments for three main reasons: (i) the direction and amplitude of our motion are very accurately known from previous measurements \citep[e.g.,][]{Kogut1993,Fixsen1996,Hinshaw2009,Planck2014,Planck2020Overview}, (ii) the dipole is a relative, rather than absolute, measurement, which provides resilience to certain systematic effects, and (iii) the dipole signal is to leading order equal to the derivative of the monopole, and so can be used to provide a prediction for monopole measurements and/or internal consistency check for an experiment targeting both the monopole and dipole \citep[see also][]{Deshpande2018}.

The intrinsic weakness of the dipole signal certainly poses a challenge, however, 21-cm experiments are not in general limited by statistical noise. For example, $\sim$ mK noise levels can be achieved in reasonable integration times $\lesssim 10^3$ hours, which is more than enough to achieve strong detections of the monopole \citep[e.g.,][]{Harker2012,Liu2013}, and is comparable to the expected dipole amplitude. \citet{Ignatov2023} showed that the dipole could potentially be detected from the ground with pre-existing monopole experiments. However, observations from space are certainly ideal given the need for $\sim$ all sky coverage, the potential for Earth's ionosphere to induce spectral distortions of greater magnitude than the 21-cm monopole \citep{Vedantham2014,Datta2014,Shen2021}, and the potential for shielding from radio frequency interference on the lunar farside \citep[hence the sustained interest in lunar observatories, e.g.,][]{Burns2012,Burns2017,FARSIDE,Pratush,Chen2019,Shi2022}.

Our focus in this paper is on what can be learned about the first stars and galaxies from a 21-cm dipole detection. \citet{Hotinli2023} recently investigated the value added to a monopole measurement from a dipole detection\footnote{Note that \citet{Hotinli2023} also considered the quadrupole, finding it to be about $\sim 100$ times weaker than the dipole. As a result, we will focus only on the dipole in this work.}, showing that the dipole can provide some modest improvement in constraints on astrophysical parameters, but can dramatically improve constraints on the foreground. Here, our approach is slightly different, but complementary, in that we focus on what can be learned from \emph{only} a dipole detection vs. \emph{only} a monopole detection, with and without priors from other astrophysical probes. Furthermore, we explore the degree to which reconstructions of the monopole from dipole measurements are model-dependent, and in so doing quantify the robustness of monopole/dipole consistency checks. Our forecasts are quite idealized in that we adopt simplistic galactic foreground models and pure radiometer noise appropriate for a given integration time, sky temperature, and bandwidth. This is to focus our attention on the information content of the dipole before considering real-world measurement challenges, given that this is a relatively unexplored topic. We defer a more detailed forecast to Anderson et al., in preparation.

In \S\ref{sec:models}, we outline our approach to modeling the 21-cm monopole and dipole in the context of high-$z$ galaxy survey results, and present the basic expectations for the dipole in this framework. In \S\ref{sec:results}, we present the results of our forecast, comparing dipole- and monopole-based constraints. We conclude in \S\ref{sec:conclusions}.  We use cosmological parameters consistent with the recent \citet{Planck2018} constraints: $\Omega_m = 0.3156$, $\Omega_b = 0.0491$, $h = 0.6726$, $n_s=0.9667$, and $\sigma_8=0.8159$.

\section{Modeling Approach} \label{sec:models}

\subsection{21-cm signals}
To model the 21-cm monopole and dipole, we use the \textsc{ares} code\footnote{\url{https://github.com/mirochaj/ares}; revision 8c8992c}. \textsc{ares} starts from cosmological initial conditions after recombination \citep{CAMB,CosmoRec}, and treats the intergalactic medium as a two-phase medium, i.e., it evolves separately the volume-filling factor of ionized bubbles, $Q$, and the properties of the mostly-neutral ``bulk IGM'' beyond. We summarize the pertinent details here and refer the interested reader to \citet{Mirocha2014} for more information on the underlying algorithm.

The 21-cm dipole\footnote{Throughout, we use `dipole' to mean `kinematic dipole.' We expect the clustering dipole to be much smaller than any kinematic dipole we consider, $\lesssim 10 \ \mu\rm{K}$, based on extrapolating the large-angular scale predictions from \citet{Zaldarriaga2004} and scaling to the pre-reionization era.} is given by \citep{Slosar2017}
\begin{equation}
    \Delta T_{\rm{dip}} = \left(\delta T_b - \frac{d\delta T_b}{d\nu} \nu \right) \frac{v_d}{c} \cos\theta \label{eq:dip}
\end{equation}
where $v_d/c \simeq 1.2 \times 10^{-3}$ is our velocity, $\cos\theta$ is the angle relative to the dipole peak, and $\delta T_b$ is the usual expression for the 21-cm monopole \citep[e.g.,][]{FurlanettoOhBriggs2006},
\begin{equation}
  \Delta T_{\rm{mon}} \equiv \delta T_b \simeq 27  \
   \overline{x}_{\rm{H}\textsc{i}} \left(1 - \frac{T_{\gamma}}{T_S} \right) \left(\frac{1 + z}{10} \right)^{1/2} \ \mathrm{mK} , \label{eq:dTb}
\end{equation}
where $1 - \overline{x}_{\rm{H}\textsc{i}} = Q + (1 - Q) x_e$ is the mean ionized fraction, broken into a fully-ionized phase with volume fraction $Q$ and bulk IGM with electron fraction $x_e$, set by the ionized fraction of hydrogen as well as helium, $T_{\gamma}$ is the background temperature, assumed here to be the CMB temperature, and $T_S$ is the spin temperature in the bulk IGM,
\begin{equation}
    T_S^{-1} = \frac{T_{\gamma}^{-1} + x_{\alpha} T_{\alpha}^{-1} + x_c T_K^{-1}}{1+x_{\alpha} + x_c} .
\end{equation}
We make the usual assumption that the temperature of the UV radiation field is equivalent to the kinetic temperature, $T_{\alpha} \approx T_K$, take collisional coupling coefficients $x_c$ from \citet{Zygelman2005}, and compute the Wouthuysen-Field coupling $x_{\alpha}$ following \citet{FurlanettoPritchard2006} \citep[see also,e.g.,][]{Hirata2006,Chuzhoy2006,Mittal2021}. Note that $x_{\alpha} \propto J_{\alpha}$, where $J_{\alpha}$ is the intensity of the Ly-$\alpha$ background.

Our model effectively neglects correlations in the density, ionized fraction, and spin temperature, i.e., we assume that the average brightness temperature is equal to the product of averages, rather than computing the product of the constituent quantities at the field level and \textit{then} averaging. This is a common approximation in the global 21-cm literature, and was recently shown to be accurate at the $\sim 10$\% level \citep{Schaeffer2024}.

We employ three different parameterizations for the 21-cm signal in order to better illustrate the information content of the dipole relative to the monopole, and to test the model-dependence of dipole vs. monopole consistency checks:
\begin{enumerate}
  \item A phenomenological model in which $J_{\alpha}$, $T_S(z)$, and $\xHI(z)$ are all given by $\tanh$ functions \citep{Harker2016}. We refer to this as the {\bf phenomenological model} or {\bf $\tanh$ model}.
  \item A physically-motivated semi-empirical model, in which $f_{\ast}$ is parameterized as a function of halo mass (and optionally redshift) in order to better fit high-$z$ rest-ultraviolet luminosity functions \citep[UVLFs;][]{Mirocha2017}. We refer to this as the double power-law or {\bf DPL $f_{\ast}$ model}  since we parameterize $f_{\ast}$ as a double power-law.
  \item An ``extended'' DPL model ({\bf DPLX}), that includes extra parameters in order to allow more flexible behaviour in the low-mass galaxy population, to be described momentarily.
\end{enumerate}
For the DPL and DPLX models, we have
  \begin{equation}
    f_{\ast} = \frac{f_{\ast,10} \ \mathcal{C}_{10}} {\left(\frac{M_h}{M_{\mathrm{p}}} \right)^{-\alphalo} + \left(\frac{M_h}{M_{\mathrm{p}}} \right)^{-\alphahi}} \label{eq:sfe_dpl}\,.
\end{equation}
where we have normalized $f_{\ast}$ to halos with $M_h=10^{10} \ M_{\odot}$ via the parameter $f_{\ast,10}$, hence the re-normalization factor $\mathcal{C}_{10}$ in our formula. We denote the peak of the double power-law $M_p$, while the low and high-mass slopes are denoted $\alphalo$ and $\alphahi$, respectively. The star formation efficiency thus requires four parameters on its own ($f_{\ast,10}$, $M_{\rm{peak}}$, $\alphalo$, $\alphahi$). Star formation is allowed to proceed in halos down to a truncation mass, implemented as a smooth exponential decline in $f_{\ast}$ at $M_h \lesssim M_{\rm{turn}}$,
\begin{equation}
    T(M_h) \equiv \left\{1 - \exp\left[-\left(\frac{M_h}{M_{\rm{turn}}} \right)^{r_{\rm{turn}}} \right] \right\} .
\end{equation}
where $r_{\rm{turn}}$ controls the sharpness of the turn-over.  This is the approach taken in \textsc{21cmfast} as well \citep[with $r_{\rm{turn}}=1$;][]{Park2019}.

We also vary the escape fraction of ionizing photons, $f_{\rm{esc}}$, and the X-ray luminosity -- SFR relation, $L_X/\rm{SFR}$, as well as two parameters that govern the low-mass behaviour of $f_{\ast}$ (see below), resulting in an eight parameter DPL model. In this work, for simplicity we do not allow for redshift evolution in any of these parameters. The model in its current form is quite flexible already, particularly with the addition of additional low-mass extensions described below, but this could be an interesting avenue for future study.

The DPL model space is fairly compact relative to older ``$f_{\rm{coll}}$ models'' \citep[see, e.g.,][]{Barkana2005,Furlanetto2006} given that the UVLFs narrow the range of allowed cosmic star formation histories. Though it could not fit an arbitrary 21-cm signal, it establishes a well-motivated target and null hypothesis to test with observations. Rejection of this null hypothesis would immediately indicate the presence of ``new'' source populations \citep[as the EDGES signal does;][]{Mirocha2019,Schauer2019,Mebane2020,Chatterjee2020}, e.g., PopIII stars, proto-quasars, a departure from the scaling relations that star-forming galaxies appear to follow, or deviations in the abundance of DM halos themselves. For example, while a turn-over in the UVLF presumably encodes the physics of feedback, an up-turn could mimic bursty star formation models, or PopIII scenarios which result in elevated star formation efficiencies in galaxies occupying low-mass halos. This is known to induce a more gradual descent into absorption in the monopole signal \citep{Mirocha2018,AhnShapiro2021,Hegde2023}, and as a result could make the dipole harder to detect. In any case, predictions for PopIII star formation span a broad range of possibilities \citep[e.g.,][]{Jaacks2018,Mebane2018,GesseyJones2022,Munoz2022,Feathers2024,Ventura2024}, so a flexible approach is warranted\footnote{Yet more flexibility could be warranted, e.g., allowing the SED of galaxies to vary as a function of mass, redshift, and/or metallicity, though for simplicity we neglect these possibilities here.}.

The point of the DPLX model is to accommodate such scenarios without invoking a specific physical model, by allowing an optional phenomenological extension to $f_{\ast}$ at faint UV magnitudes, i.e.,
\begin{equation}
  f_{\ast} \rightarrow f_{\ast} S(M_h) .
\end{equation}
We follow \citet{Schneider2021}, who defined this `small scale' function as
\begin{equation}
  S(M_h) \equiv \left[1 + \left(\frac{M_c}{M_h} \right)^{\gamma_1} \right]^{\gamma_2}
\end{equation}
which allows a suppression \textit{or} boost at $M_h \lesssim M_c$. Note that at corners of parameter space corresponding to a boost, there is a possibility that $f_{\ast}$ diverges as $M_h \rightarrow 0$. We impose $f_{\ast} \leq 1$, however, in practice our truncation function $T(M_h)$ takes over and prevents this from occurring.

Note that our fiducial scenario (model A) assumes $S=1$, and so we will presumably achieve better constraints on $M_{\rm{turn}}$ if one assumes $S=1$ in the fitting as well (i.e., if one uses the DPL instead of the DPLX model). However, focusing only on $M_{\rm{turn}}$ in a fit with $S$ allowed to vary could be misleading, since new degeneracies with $M_c$, $\gamma_1$, and $\gamma_2$ could achieve the same turn-over properties in, e.g., UV magnitude, but very different values of $M_{\rm{turn}}$. As a result, in \S\ref{sec:results} we will focus on the recovery of $M_{\rm{turn}}$ and $r_{\rm{turn}}$ for the DPL model, but for the DPLX model we will focus on the recovery of derived quantities like the UVLF turn-over (in $M_{\rm{UV}}$ and $\phi[M_{\rm{UV}}])$.

Finally, note that for the DPL and DPLX models, we do not separately vary the strength of the non-ionizing UV emission via $N_{\rm{LW}}$. Instead, the relative strength of ionizing and non-ionizing emission is determined self-consistently by the stellar population synthesis (SPS) models we employ, \textsc{bpass} \citep{Eldridge2009,Eldridge2017}, for a metallicity of $Z=0.004$. Note that the stellar population assumptions will be degenerate with $f_{\ast,10}$, so we do not vary them separately -- constraints on $f_{\ast,10}$ should thus be interpreted with caution\footnote{Note that, e.g., stellar metallicity does not affect the non-ionizing and ionizing output of galaxies identically, so introducing this as a free parameter could in principle introduce new behaviour, but this is likely a small effect \citep[see, e.g., Appendix A in][]{Mirocha2017}.}.

\begin{figure*}
\begin{center}
\includegraphics[width=0.98\textwidth]{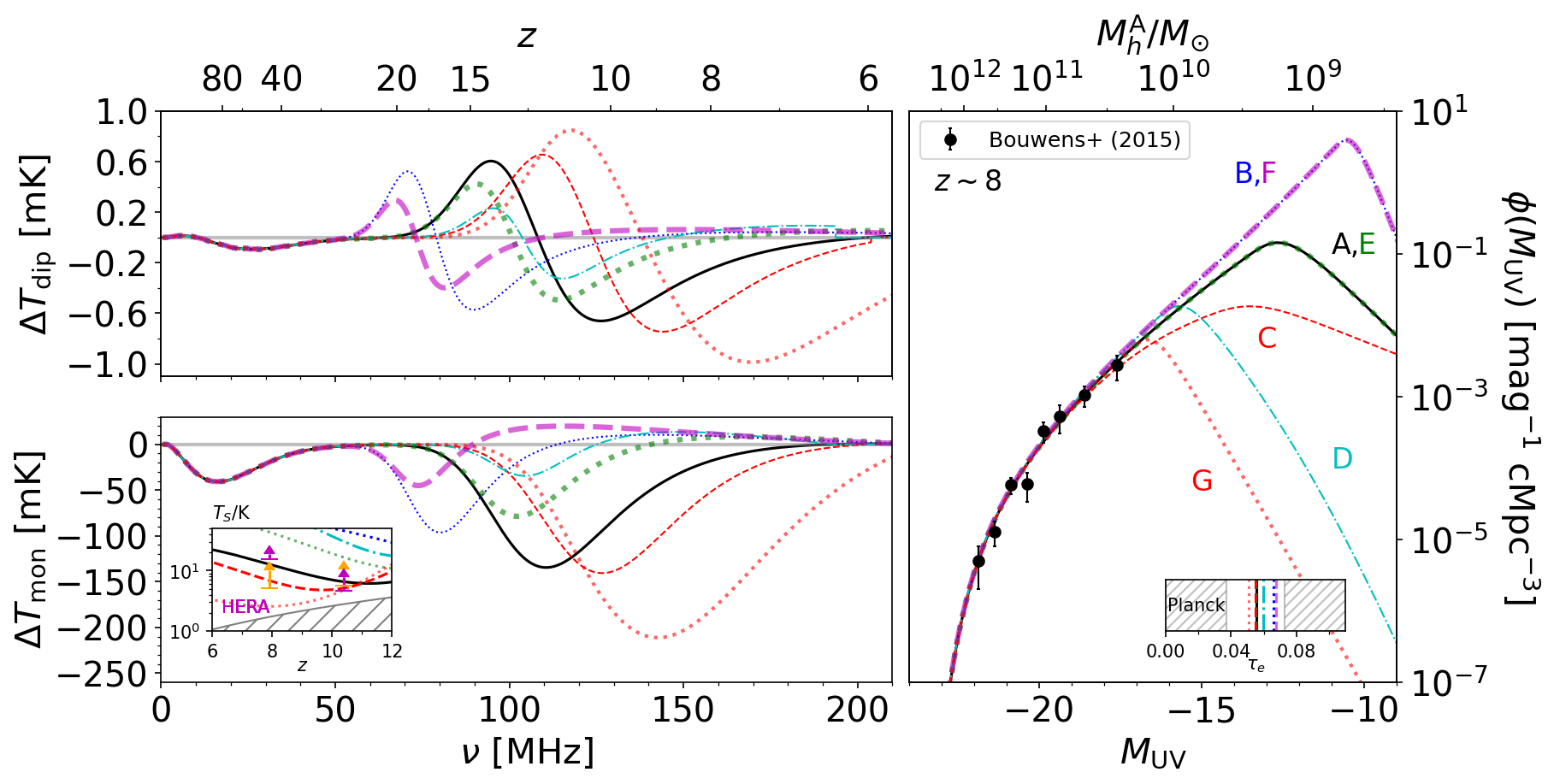}
\caption{{\bf Models explored in this work.} Each model for the 21-cm dipole (top left) and monopole (bottom left) are anchored to UVLFs (right) from \citet{Bouwens2015}. Black solid curves indicate our fiducial scenario ("model A"). Additional curves indicate scenarios indistinguishable via UVLFs (right) and CMB optical depth constraints from \citet{Planck2018} (inset, lower right), achieved by changing the behaviour at the faint end. Model C (dashed red) is in mild tension with the latest 21-cm power spectrum limits from HERA \citep{HERA2023} (inset, lower left), while model G (dotted red) is strongly disfavoured by HERA. Note that the $M_h$ axis along the top of the right panel is model-dependent, and corresponds to model A only.}
\label{fig:story}
\end{center}
\end{figure*}

Example realizations of the DPL/DPLX model are shown in Fig. \ref{fig:story}. Each case matches $z \gtrsim 6$ UVLFs by construction but has different behaviour at the faint end of the UVLF. The faint-end differences drives qualitatively different behaviour in the 21-cm signals, all of which are consistent with pre-existing constraints, at least roughly. In each case, we effectively assume high-mass X-ray binaries are the dominant heat sources, because we adopt a multi-colour disk spectrum for 10 $M_{\odot}$ black holes \citep{Mitsuda1984} which results in a relatively hard X-ray spectrum and so inefficient heating \citep{Mirocha2014}, similar to the effects of a Cygnux X-1 template \citep{Fialkov2014HMXBs}, which drives later features than earlier generations of models. We vary the relationship between star formation and X-ray photon production, starting with a fiducial value of $L_X/\rm{SFR}$ of $2.6 \times 10^{39} \ \rm{erg} \ \rm{s}^{-1} \ (M_{\odot}/\rm{yr})^{-1}$ in the 0.5-8 keV band, which is representative of local star-forming galaxies \citep{Mineo2012}, but include models with values as high as $10^{42}$ as well. Note that the default value in, e.g., \textsc{21cmfast} is higher, $\sim 10^{40.5}$, as is expected of low-metallicity systems \citep[e.g.,][]{Fragos2013,Brorby2016,Lehmer2022}, and so generally results in a weaker absorption signal than our fiducial model (model A). Finally, we redden the intrinsic spectrum of galaxies with an optical depth determined by a characteristic column density $\log_{10} N_{\rm{H} \textsc{i}} = 21$, which is consistent with simulations \citep{Das2017}. In principle 21-cm fluctuations can break this degeneracy \citep[see, e.g.,][]{Pacucci2014}, but for monopole and/or dipole measurements this will be much more difficult. As a result, we keep $\log_{10} N_{\rm{H} \textsc{i}}$ fixed for simplicity.

\begin{deluxetable}{lccccccc}
\tabletypesize{\scriptsize}
\tablecolumns{8}
\tablecaption{Models explored in this work
\label{tab:params}}
\tablehead{\colhead{parameter} & \colhead{A} &  \colhead{B} &  \colhead{C} &  \colhead{D} &  \colhead{E} &  \colhead{F} &  \colhead{G}}
\startdata
$L_X/\rm{SFR}$ & $10^{40.7}$   & $10^{41}$ & $10^{40.7}$    & $10^{42}$ &  $10^{41.5}$ & $10^{41.7}$ & $10^{40}$ \\
$f_{\rm{esc}}$ & 0.15          & 0.1    & 0.25 & 0.25 & 0.15 & 0.1 & 0.2 \\
$M_{\rm{turn}}/M_{\odot}$      & $10^9$ & $10^8$ & $10^{9.5}$ & $10^{9.7}$  & $10^9$  & $10^8$ & $10^{10}$  \\
$r_{\rm{turn}}$ & 3            & 3.5    & 3 & 3.5  & 3  & 4 & 4.5   \\
\hline
$M_{\rm{crit}}/M_{\odot}$      & n/a    & $10^9$ & $10^{9.5}$ & $10^{9}$  & n/a & $10^9$ & n/a  \\
$\gamma_1$ & n/a & -0.75 & -1.5 & -0.75 & n/a   & -0.75 & n/a \\
$\gamma_2$ & n/a & 1     & -2   & -1    & n/a   &  1 & n/a \\
\enddata
\tablecomments{Key parameters for DPL/DPLX models shown in Fig. \ref{fig:story}. All models adopt the same double power law parameters, $f_{\ast,10}=0.02$, $M_{\rm{peak}} = 2 \times 10^{11} \ M_{\odot}$, $\alpha_{\ast,\rm{lo}} = 0.49$, $\alpha_{\ast,\rm{hi}}=-0.61$, and the same X-ray spectrum: a multi-colour disk model for a 10 $M_{\odot}$ black hole with intrinsic hydrogen absorbing column of $\log_{10} N_{\rm{HI}} /\rm{cm}^{-2} = 21$. Variations in $M_{\rm{turn}}$, $r_{\rm{turn}}$, $M_{\rm{crit}}$, and the $\gamma$ parameters are chosen to span a wide range of qualitatively different possibilities for the UVLF faint-end. Note that models A, E, and G take $S=T=1$, hence the 'not applicable' labels in the final three rows. Models with a faint-end 'up-turn' are meant to mimic scenarios with elevated star formation efficiencies or burstiness at low mass.}
\end{deluxetable}

\subsection{Prior volume} \label{sec:priors}
The models shown in Fig. \ref{fig:story} agree with high-$z$ UVLFs by construction, with values for remaining free parameters like $f_{\rm{esc}}$ tuned to provide good agreement with $\tau_e$ from \textit{Planck} and $L_X/\rm{SFR}$ set to values that (mostly) jive with the latest 21-cm power spectrum limits from HERA \citep{HERA2023}. Here, we take a more thorough look at the prior volume for 21-cm monopole and dipole measurements.

To do this, we vary all 11 of the DPLX model's free parameters subjected to the following constraints: (i) the CMB optical depth $\tau = 0.055 \pm 0.009$ from \citet{Planck2018}, (ii) Constraints on the end of reionization from the Ly$\alpha$ forest -- we conservatively assume that reionization must be complete ($\xHI \leq 1$\%) by $z=5.3$, in line with recent measurements \citep[e.g.,][]{Bosman2022}, and (iii) UVLFs from \citet{Bouwens2015} at $z \sim 6-8$. The last is particularly important to some of our results -- given that the 21-cm monopole and dipole probe only the volume-averaged emissivity of galaxies, one would not expect to be able to constrain the shape of $f_{\ast}$ in detail without including UVLFs in the likelihood (see, e.g., Fig. 6 in \citet{DorigoJones2023}). Our approach to reionization priors is conservative in that we neglect pre-existing constraints on the detailed redshift evolution of the neutral fraction from, e.g., Ly-$\alpha$ emitters \citep[e.g.][]{Mason2018}, or quasar damping wings \citep[e.g.,][]{Davies2018,Greig2019}.

We will also compare to lower limits on the spin temperature of the $z \sim 8$ IGM from HERA \citep{HERA2023} and limits on the unresolved fraction of the cosmic X-ray background \citep{Hickox2006,Lehmer2012}, but we do not actually impose either of the latter two constraints as priors here or in subsequent forecasting. We also neglect all pre-existing constraints on the global 21-cm signal \citep[see, e.g.,][for examples]{Monsalve2017,Singh2017} in order to remain agnostic about possibilities for the dipole. Joint constraints from monopole and power spectrum measurements are quickly becoming interesting \citep{Pochinda2023,Bevins2024}, though for simplicity we defer such considerations for the dipole to future work.

\begin{figure*}
\begin{center}
\includegraphics[width=0.98\textwidth]{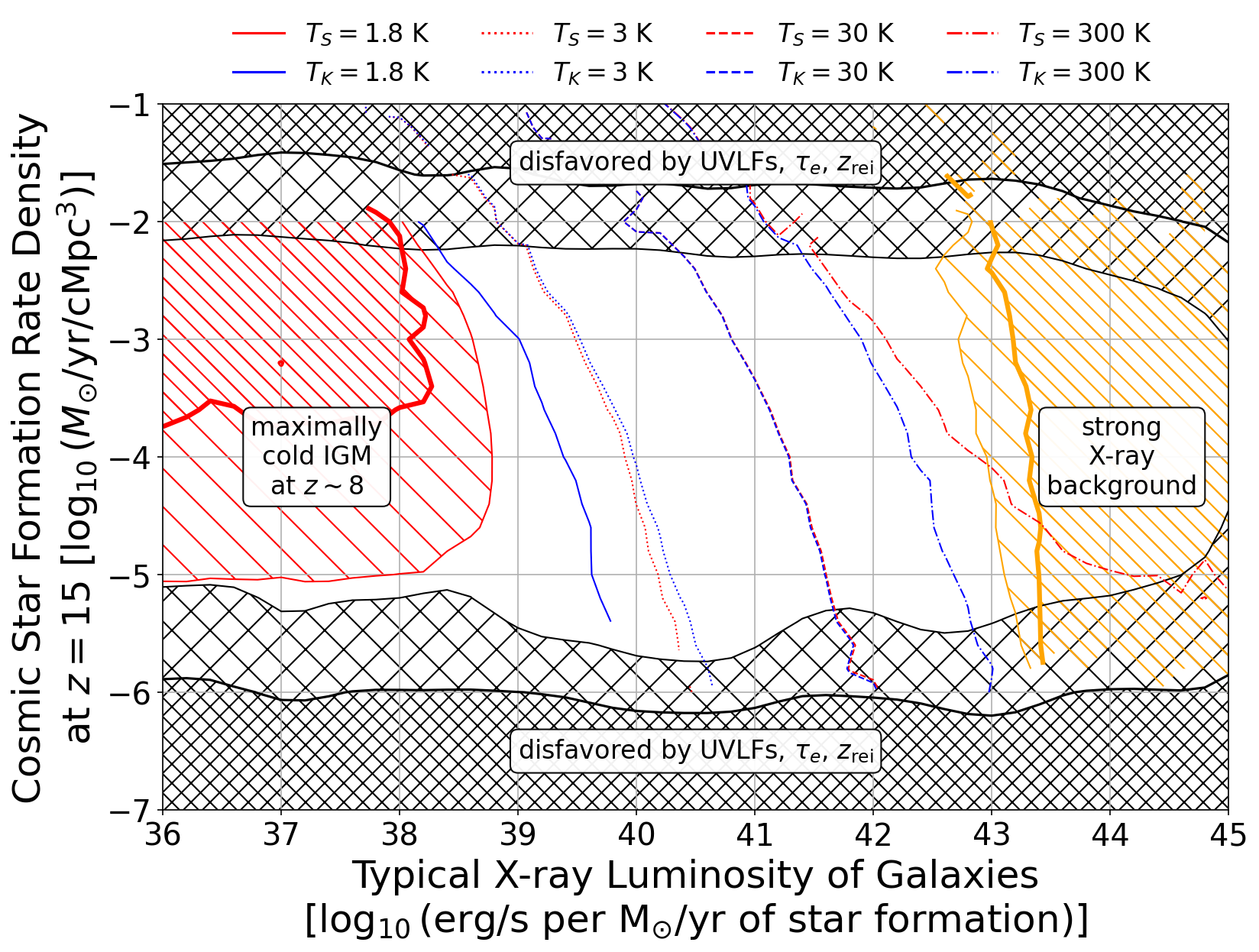}
\caption{{\bf Condensed discovery space for 21-cm monopole and dipole measurements.} For illustrative purposes, here we focus on a 2-D slice of parameter space in which 21-cm measurements can provide powerful constraints on the first galaxies. As a barometer for high-$z$ global star formation activity, we focus on the $z=15$ cosmic SFRD ($y$ axis), while the normalization of the $L_X/\rm{SFR}$ relation is shown on the $x$-axis as a black hole activity indicator. Reionization constraints \citep{Planck2018,Bosman2022} and UVLFs \citep{Bouwens2015} disfavour the very top and bottom regions of this diagram (cross-hatching fills the $2$ and $3\sigma$ disfavoured regions). Colored contours along the right hand side represent constraints on the unresolved fraction of the cosmic X-ray background \citep[in the soft band from \textit{Chandra};][]{Lehmer2012}, which disfavor very large values of $L_X/\rm{SFR}$. Indicated along the left is a region of parameter space in which models produce effectively no heating of the $z \sim 8$ IGM, a scenario which is now disfavoured by 21-cm power spectrum experiments \citep{HERA2022Obs,MWA2020Obs}. Note that the X-ray and 21-cm contours are not confidence intervals; instead, they enclose regions in which \textit{all} models violate the given constraint (dense cross-hatching) and regions where only \textit{some} models remain consistent with these constraints (sparser cross-hatching). Finally, we also draw contours at fixed $z=8$ spin and kinetic temperatures of 3, 30, and 300 K (mean value in each pixel) as indicated along the top of the figure, as a rough guide for how future constraints map to this space. Also apparent are the regimes in the lower left and right corners where the spin temperature is not fully coupled to the kinetic temperature. At $z \sim 10$, the lower boundary of these `weak coupling' regions shift upward by roughly an order of magnitude. We use 0.2 dex wide pixels in each dimension to determine the number of models in each ($\sim 50-100$ on average), and smooth with a gaussian kernel for contours. Note that the boundaries of these regions are subject to assumptions about the SEDs of galaxies, see text for details.}
\label{fig:discovery_space}
\end{center}
\end{figure*}

Despite the priors listed above, there is still a considerable range of possibilities, particularly for the faint-end of the galaxy UVLF. The wide range of viable possibilities in the UVLF imply a wide range of possibilities for the cosmic SFRD, which provides a nice way to compress the many parameters controlling $f_{\ast}$ into a single, albeit redshift-dependent, quantity. As a result, in Fig. \ref{fig:discovery_space}, we show how the SFRD -- $L_X/\rm{SFR}$ space is constrained by current observations. This plot is constructed from the results of a prior-only fit, and is composed of $\sim 570,000$ models. The cross-hatched region along the top is disfavored by UVLFs and reionization constraints, as models in this space produce too many bright galaxies relative to measured UVLFs and/or a reionization epoch that ends too quickly to remain in agreement with CMB $\tau_e$ or Ly-$\alpha$ forest constraints on the end of reionization. In the top-right corner of this space, we see that the disfavoured region grows slightly for very large $L_X/\rm{SFR}$ values, indicating that it is X-ray sources that are responsible for finishing reionization too quickly. The cross-hatched region along the bottom is disfavoured by the same set of observations, but in this case because reionization occurs too late and/or there are too few high-$z$ galaxies relative to UVLFs.

We also show the region of parameter space disfavoured by constraints on the soft X-ray background \citep[e.g.][]{Hickox2006,Lehmer2012}, which unsurprisingly corresponds to large values of $L_X$/\rm{SFR}. We adopt a 0.5-2 keV X-ray background intensity of $1.96 \ \times \ 10^{-12} \ \rm{erg} \ \rm{s}^{-1} \ \rm{cm}^{-2} \ \rm{deg}^2$, which is the total intensity $8.15 \ \times 10^{-12} \ \rm{erg} \ \rm{s}^{-1} \ \rm{cm}^{-2} \ \rm{deg}^2$ from \citet{Lehmer2012} times their best-fit unresolved fraction of 24\%. The denser hatching indicates the region of parameter space in which \textit{all} of our models violate this constraint, whereas the less dense hatching is more generous, including regions of parameter space in which some -- but not all -- models are disfavoured by the X-ray background. Note that analogous constraints on the diffuse radio background \citep{Fixsen2011,Dowell2018} provide a useful diagnostic for excess radio background models \citep[see, e.g.][]{EwallWice2018,Fialkov2019}, though in this work we take $T_R = T_{\rm{CMB}}$.

A similar cross-hatched region borders the left edge of the plot, and indicates scenarios in which the $z \sim 8$ IGM is cold during reionization. By ``cold,'' here we mean that the bulk IGM (the portion of the IGM that is mostly neutral) is completely unheated, resulting in \textit{spin} temperatures equal to the theoretical minimum in $\Lambda$CDM, $T_S \simeq 1.8$ K at $z=8$. Such scenarios are now disfavoured, so we also include contours corresponding to mean IGM spin temperatures of 3, 10, 100, and 300 K to roughly indicate how current and future 21-cm power spectrum limits map to this parameter space. Note that there is a gap in the lower left corner of the plot, where the SFRD is low but the IGM is apparently not maximally cold. This region is populated by models that \textit{are} maximally cold, but lack a sufficiently strong Ly-$\alpha$ background to couple $T_S \rightarrow T_K$ (compare to blue contours of same linestyle), and so cannot yet be fully ruled out by 21-cm measurements, at least when limiting analyses to this single band. Most of this region is disfavoured at $\sim$ 2$\sigma$ by our UVLF and EoR priors, but should be scrutinized more carefully in future multi-epoch 21-cm analyses. At $z\sim 10$, this `weak coupling' region is roughly an order of magnitude larger in the SFRD dimension.

As is always the case for 2D representations of $>2$D parameter spaces, Fig. \ref{fig:discovery_space} does not tell the whole story. For example, our assumptions for the X-ray SED of sources -- kept fixed here -- surely affects the precise location of the red and orange cross-hatched regions. A softer SED achieved, e.g., by decreasing $\log_{10} N_{\rm{H} \textsc{i}}$, would result in more efficient heating per unit star formation, and so shrink the `maximally cold IGM' region. Similarly, a harder X-ray spectrum would cause more tension with the X-ray background, and allow the orange `strong X-ray background' region to grow. For the X-ray background, we have also used the flux generated by our model for all sources at redshifts higher than $z_{\rm{min}} = 6$. Of course in reality, more aggressive removal of sources can further reduce the unresolved fraction, perhaps from $\sim 24\%$ \citep{Lehmer2012} to $\sim 3$\% \citep{Cappelluti2012}, which would allow the orange region to grow. A self-consistent treatment of the 21-cm background and X-ray number counts would provide a more careful accounting of the unresolved fraction and $z_{\rm{min}}$, though is beyond of the scope of this work.

\begin{figure}
\begin{center}
\includegraphics[width=0.49\textwidth]{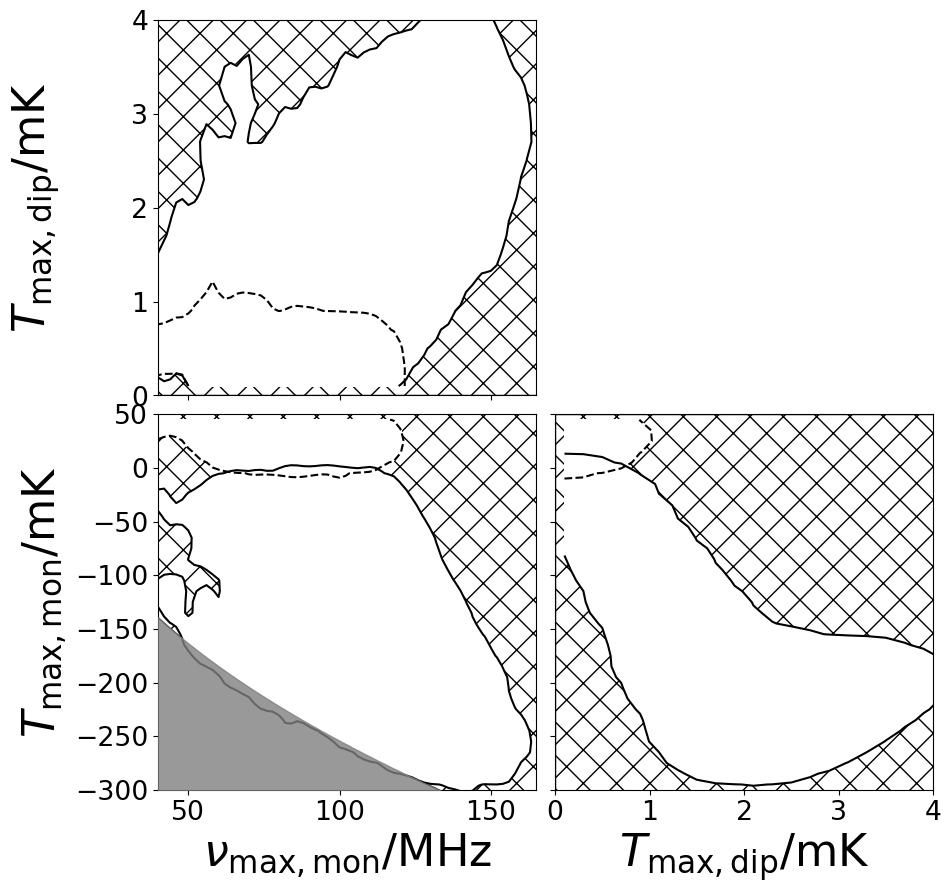}
\caption{{\bf Expected monopole and dipole characteristics across the DPL model's prior volume.} Here, as in Fig. \ref{fig:discovery_space}, the black cross-hatched region indicates parts of parameter space disfavoured at $\geq 3\sigma$ by UVLFs, $\tau_e$, and $z_{\rm{rei}}$. We separate models for which the monopole peaks in absorption vs. emission ($\lesssim 10\%$ of prior volume, hence the secondary mode with $T_{\rm{max,mon}} > 0$ in each panel (dashed contours). Note that the dipole amplitude is measured peak to trough, whereas the quoted monopole amplitude is the amplitude of the absorption minimum or emission maximum, whichever is stronger. The dark gray region in the lower left corner indicates the brightness temperature as a function of frequency in $\Lambda$CDM with full Wouthuysen-Field coupling and no X-ray heating. Regions disfavoured by 21-cm power spectrum and X-ray background constraints are not shown here, as they `pile-up' along the edge of the available parameter space where the 21-cm monopole and dipole are strongest.}
\label{fig:pvol}
\end{center}
\end{figure}

Next, in Fig. \ref{fig:pvol}, we show how our priors map to the space of the frequency and peak amplitude of the monopole and dipole. Each panel shows a different cut through the joint distribution of the peak monopole frequency $\nu_{\rm{max,mon}}$, peak monopole amplitude $T_{\rm{max,mon}}$, and peak-to-trough dipole amplitude, $T_{\rm{max,dip}}$. Clearly, stronger monopole absorption signals result in stronger dipole signals (lower right panel), which generally correspond to late features $\nu \gtrsim 110$ MHz (left column). These are precisely the kinds of scenarios that 21-cm power spectrum limits are beginning to rule out \citep{HERA2022Theory,HERA2023,MWA2020Obs}. Models disfavoured by 21-cm and X-ray background constraints are not shown here, as they `pile up' at the boundaries of the prior volume. For example, because strong X-ray backgrounds drive $T_S \gg T_{\rm{CMB}}$, the models lining the right edge of the $L_X/\rm{SFR}$ parameter space all inhabit a narrow sliver of ($\nu_{\rm{max,mon}}$, $T_{\rm{max,mon}}$) space. Similarly, cold IGM models pile up at the edge of the gray shaded boundary in the lower left corner of Fig. \ref{fig:pvol}.

Now, we proceed to the details of our forecasting approach, including our treatment of the foreground, mock experimental uncertainties, and sampling of the parameter space.

\subsection{Foregrounds}
We take a very simple approach to foregrounds in this work in order to establish the best-case scenario for dipole inference, in which measurements are limited by statistical (radiometer) noise only\footnote{This is clearly an oversimplification. For example, removal of the galactic foreground requires exquisite knowledge of the system, particularly the chromaticity of the beam. If mis-modeled or mis-characterized, residual spectral structures can easily prevent detection of the cosmological signal \citep[e.g.,][]{Mahesh2021,Tauscher2020MAA,Tauscher2021,Hibbard2023,Sims2023,Cumner2023,Agrawal2024}.},
\begin{equation}
  \sigma_{\nu} \propto \frac{T_{\rm{sky}}}{\sqrt{t_{\rm{int}} \Delta \nu}}
\end{equation}
where $t_{\rm{int}}$ is the integration time, $\Delta \nu$ the channel width, and $T_{\rm{sky}}$ is the sky temperature as a function of frequency. We assume a simple power-law foreground spectrum,
\begin{equation}
  T_{\rm{sky}} = T_{75} \ \rm{K} \ \left(\frac{\nu}{75 \ \rm{MHz}}\right)^{\beta}
\end{equation}
with $T_{75} = 1700 \ \rm{K}$ and $\beta=-2.59$, consistent with the latest measurements directed away from the galactic plane \citep{Mozdzen2019} and extrapolations of maps at higher frequencies \citep[e.g.][]{Haslam1982,Guzman2011}.

\begin{figure}
\begin{center}
\includegraphics[width=0.49\textwidth]{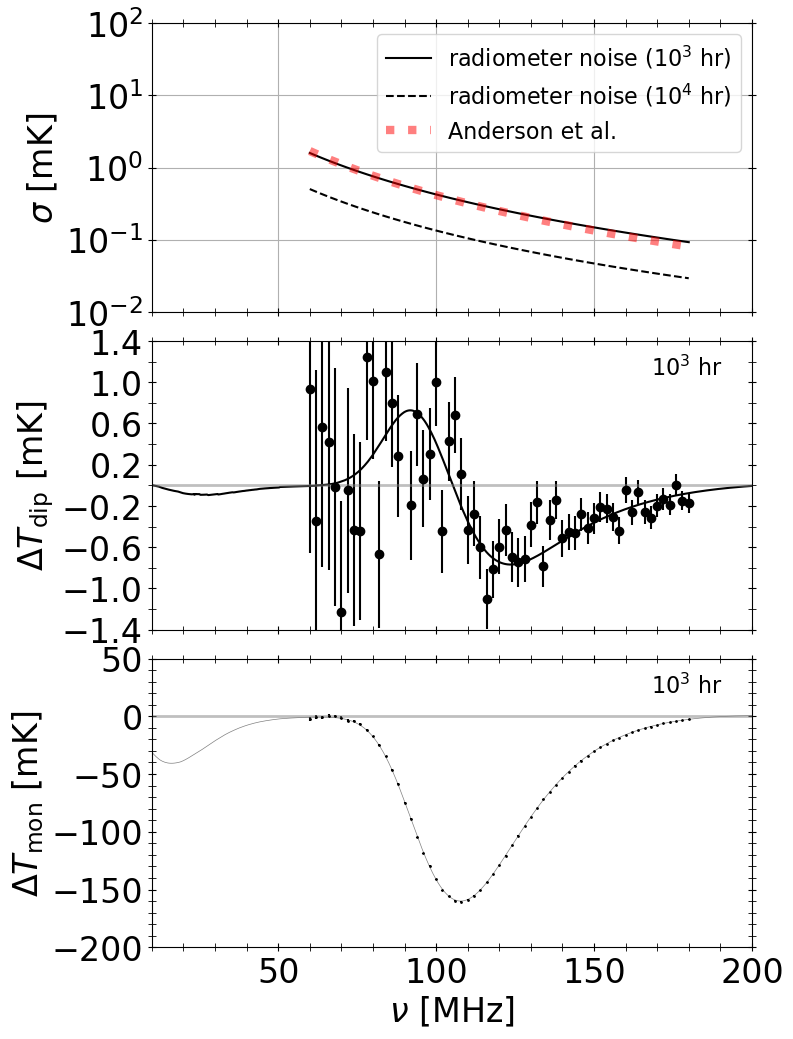}
\caption{{\bf Example mocks for input model A.} In the top row, we show the noise level for 1000 and $10^4$ hour integrations under the assumption of radiometer noise only (solid, dashed black) compared to a simulated noise realization from Anderson et al., in prep. (dotted red; see text for details). The 21-cm dipole and monopole are shown in the middle and bottom row, respectively. Uncertainties on the dipole are much more significant relative to the signal amplitude, though the integrated signal to noise is still significant. Note that for visual clarity, we plot points every 2 MHz, though the errorbars themselves are computed assuming channels 1 MHz wide.}
\label{fig:mocks}
\end{center}
\end{figure}

For a canonical 1000 hour integration, typical for monopole forecasts, these choices yield thermal noise levels of $\sim 1$ mK and below (depending on frequency). Here, we will consider a 1000 hour integration with a spectral resolution of 1 MHz, as shown in Fig. \ref{fig:mocks}, which results in thermal noise levels of $\simeq 0.4$ mK at 100 MHz. For model A, the cumulative signal-to-noise ratio over the 60-180 MHz band is $\sim 19$ for the dipole. This yields an effectively perfect measurement of the monopole. Using pure radiometer noise like this for a dipole measurement effectively assumes an idealized scan strategy, in which one alternates between measurements of the dipole maximum and minimum on the sky. One might expect this approach to yield a factor of two boost in the dipole amplitude relative to what we plotted in Fig. \ref{fig:story}, which took $\cos\theta=1$ (i.e., assumed the spectrum at the exact position of the dipole maximum). However, an additional factor of two boost in the noise on the difference spectrum cancels, resulting in no net change in the signal amplitude or its uncertainties. In practice, the non-negligible width of a realistic beam pattern would also dilute the dipole signal amplitude, i.e., we cannot just difference the signal at the dipole's exact maximum and minimum on the sky.

Given the many potential non-idealities involved in a realistic dipole experiment, we show for reference in Fig. \ref{fig:mocks} an example noise realization from Anderson et al., in prep., which includes a realistic scan strategy, beam power pattern, calibration scheme, and signal loss associated with the removal of the foreground, which is assumed to be constrained empirically \citep[as in][]{Switzer2014}. This curve assumed 8000 hours of integration, resulting in $\sigma(\nu)$ values resembling our highly-idealized 1000 hour integration. We defer an analysis of the impact of signal loss associated with foreground removal to Anderson et al., in prep.

\subsection{MCMC sampling}
In what follows, all mock parameter constraints were obtained via MCMC sampling with \textsc{emcee} \citep{ForemanMackey2013}.

Our likelihood function is simply
\begin{equation}
  \log \mathcal{L} \propto \sum_i \left[-\frac{1}{2} \frac{\left(m_i(\Theta)-d_i\right)^2}{\sigma_i^2} \right]
\end{equation}
where the index $i$ represents spectral channels, the data vector $d_i$ then represents the dipole and/or monopole measurement in channel $i$, and $\sigma_i$ is the uncertainty of data point $i$. The quantity $m_i$ is our model prediction for the dipole and/or monopole given parameters $\Theta$ (not to be confused with dipole direction $\theta$). Our priors are enumerated in \S\ref{sec:priors}.
\section{Results} \label{sec:results}
\begin{figure*}
\begin{center}
\includegraphics[width=0.98\textwidth]{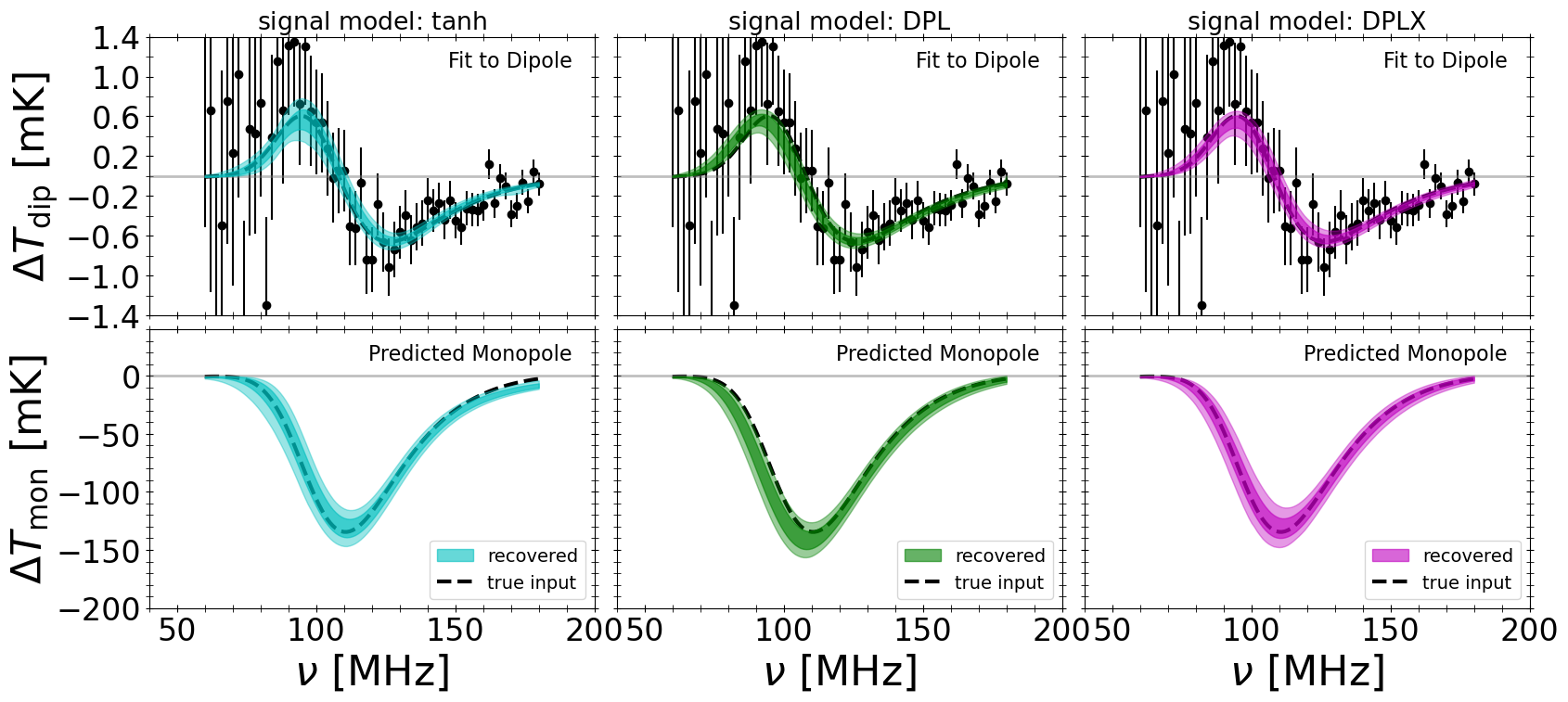}
\caption{{\bf
Measurements of the dipole provide a robust and relatively model-independent consistency check on monopole measurements.} In the top row, we show our fits to a mock dipole signal generated with the DPL model (model A; same in each panel), and in the bottom row, we show predictions for the monopole in each case. Results in each column are obtained using a different signal model to fit the mock ($\tanh$, DPL, and DPLX, from left to right). In each case, we recover the position of the absorption minimum to an accuracy of better than $\sim 20$ mK (1-$\sigma$), and find good agreement across the band as well. Note that, as in Fig. \ref{fig:mocks}, we have thinned out the number of data points plotted by 2x for visual clarity.}
\label{fig:consistency_check}
\end{center}
\end{figure*}

We begin by examining the potential of the dipole as a consistency check on monopole measurements. To do this, we fit all three of our astrophysical models -- the $\tanh$, DPL, and DPLX models -- to the same mock dipole signal (DPL model; A), to determine the degree to which the reconstructed monopole varies from model to model. The results of this exercise are shown in Fig. \ref{fig:consistency_check}. Columns correspond to the three different signal models, with the fitted dipole results shown in the top row and the reconstructed monopole in the bottom row. At a glance, it is clear that each model can provide a satisfactory fit to the dipole, despite differences in the underlying model parameterizations\footnote{Note that we also experimented with simpler four-parameter models that relate cosmic photon production to the rate of collapse of matter into dark matter halos \citep[see, e.g.,][]{Barkana2005,Furlanetto2006,Pritchard2010,Mirocha2015}. We found that some mock realizations' shapes cannot be well matched, and convergence tends to be quite slow as a result. Because of this, and the declining use of these models in the literature, we decided not to include them in our final analyses.}. Similarly, the reconstructed monopole contains the true input mock at the 68\% confidence level. The detailed shape of the reconstructed monopole does vary from model to model, particularly for the $\tanh$ model. This is because the $\tanh$ model is phenomenological, with the Lyman-$\alpha$ background, thermal history, and ionization history treated completely independently. As a result, one can (for example) recover the absorption signal well but obtain a broader range of possibilities at higher frequencies during reionization. In contrast, the more physical models generally yield a tight prediction for the monopole at high frequencies, because any astrophysical scenario that has enough heating to drive an absorption peak has enough star formation to finish reionization by $z \sim 6$.

\begin{figure}
\begin{center}
\includegraphics[width=0.49\textwidth]{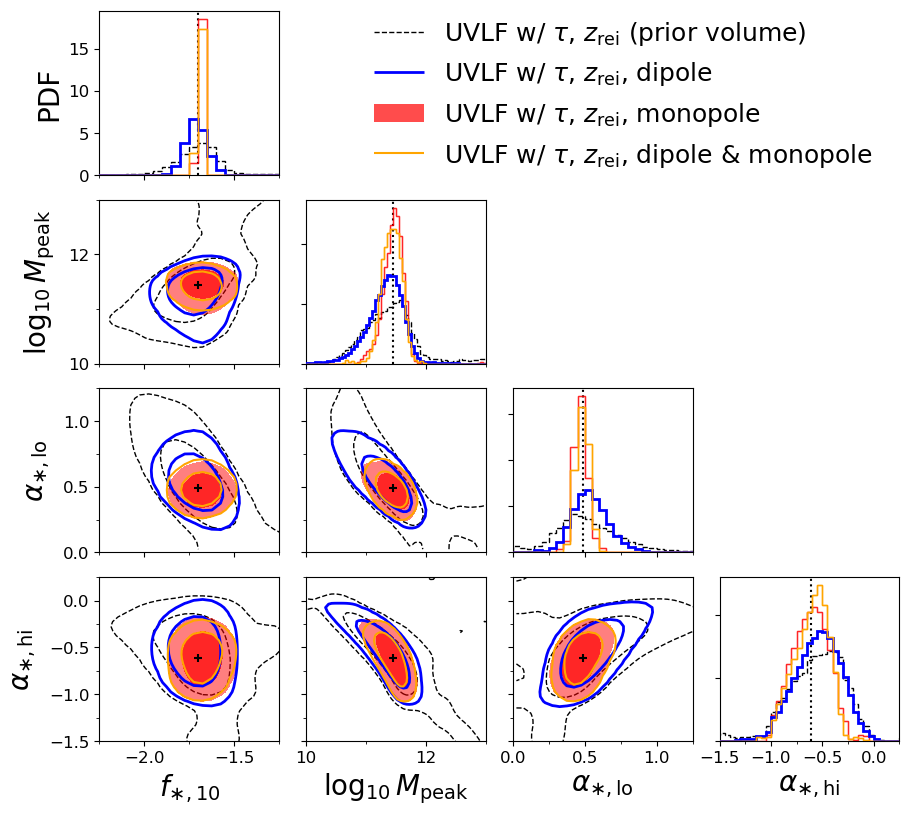}
\caption{{\bf Forecast for $f_{\ast}$ parameters.} Dashed black contours reflect the prior volume, defined by pre-existing constraints on UVLFs \citep{Bouwens2015}, Ly-$\alpha$ forest \citep{Bosman2022}, and the CMB optical depth \citep{Planck2018}. Open blue contours indicate constraints obtained with the addition of only a dipole measurement, while filled red contours show constraints obtained with the addition of only the monopole. Open orange contours denote the constraints possible with both a dipole and monopole measurement. Crosses indicate the true input values assumed in the mock (model A).}
\label{fig:constraints_21cm_sfe}
\end{center}
\end{figure}

Next, we proceed to forecast constraints on astrophysical parameters, once again focusing on the recovery of model A with the DPL model (i.e., we use the same model to fit the signal as was used to generate the mock). Starting with the $f_{\ast}$ parameters, in Fig. \ref{fig:constraints_21cm_sfe} we see that the monopole (red) provides a non-trivial improvement in the constraints on all four $f_{\ast}$ parameters relative to the prior (dashed black). The dipole constraints (blue) also exhibit improvement over the prior, though to a lesser degree than the monopole constraints. The joint constraints that incorporate both the monopole and dipole in the likelihood are nearly indistinguishable from the monopole constraints.

\begin{figure*}
\begin{center}
\includegraphics[width=0.98\textwidth]{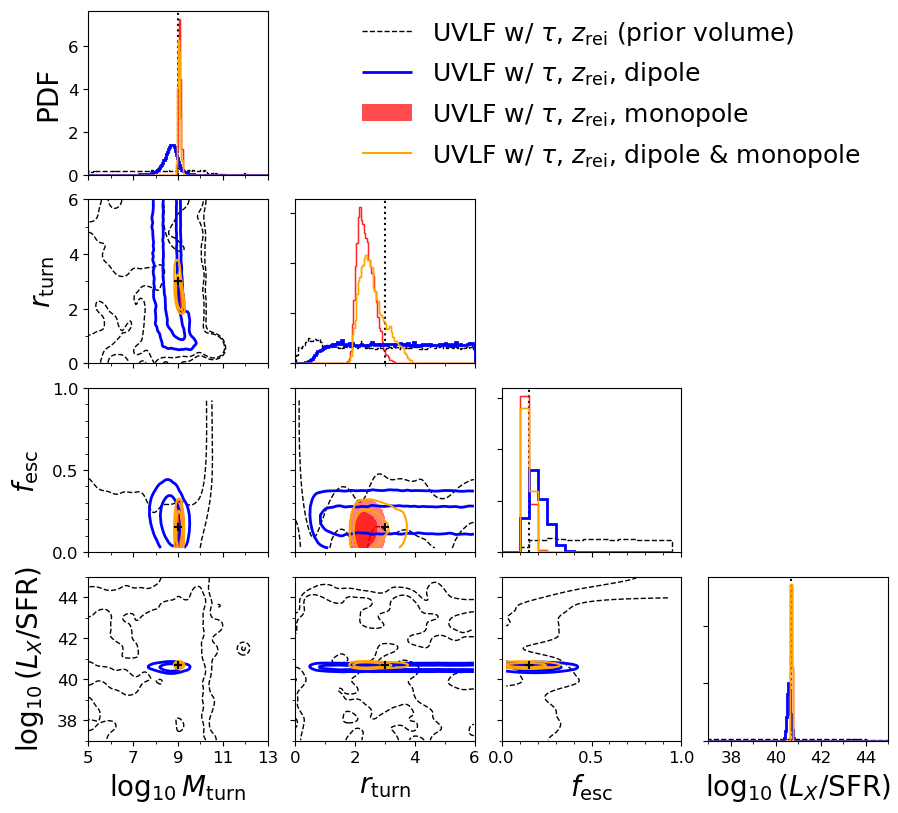}
\caption{{\bf Dipole constraints on parameters governing the faint-end UVLF turn-over and efficiency of UV and X-ray photon production are competitive with monopole constraints.} Linestyle and colour conventions are identical to Fig. \ref{fig:constraints_21cm_sfe}.}
\label{fig:constraints_21cm_other}
\end{center}
\end{figure*}

In Fig. \ref{fig:constraints_21cm_other}, we turn our attention to the remaining parameters, which govern the faint end of the UVLF and efficiency of ionizing and X-ray photon production. The prior volume here is broad, given that many combinations of, e.g., $M_{\rm{turn}}$ and $f_{\rm{esc}}$ can satisfy pre-existing constraints on reionization. Both the monopole and dipole dramatically reduce the landscape of possibilities. The monopole still outperforms the dipole, but the dipole provides powerful constraints on three of these four key parameters all on its own. The most noticeable shortcoming is in the $r_{\rm{turn}}$ dimension, which controls the `sharpness' of the turn-over in the UVLF. Conceptually, the explanation is simple: once the UVLF turns over at magnitudes corresponding to $M_{\rm{turn}}$, the cumulative number of photons coming from $M_h \lesssim M_{\rm{turn}}$ drops exponentially, meaning changes to $M_{\rm{turn}}$ will have a much bigger impact on the signal than $r_{\rm{turn}}$. The fact that monopole measurements can bound this quantity at all speaks to the power of the detailed shape of the monopole, and as a result its ability to put perhaps surprisingly informative constraints on low-mass halos \citep[see also][]{Hibbard2022}. Although the 21-cm monopole is more statistically powerful and constraining, the dipole may be more robust to a range of systematic concerns, as discussed in \S\ref{sec:intro} and in Anderson et al. (in prep). It is thus encouraging that dipole measurements alone may provide interesting parameter bounds and astrophysical insights.

\begin{figure}
\begin{center}
\includegraphics[width=0.49\textwidth]{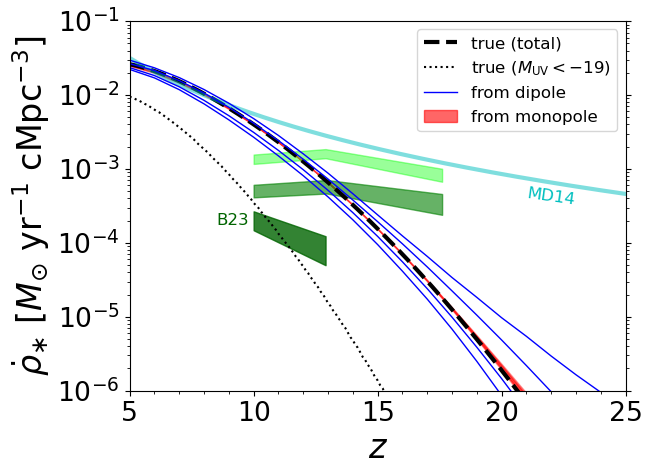}
\caption{{\bf Mock constraints on the cosmic SFRD from the 21-cm dipole and monopole.} The true input (dashed) is recovered by the dipole (open blue contours) to within $\sim 0.1$ dex at $z \sim 10-15$, with broadening contours at $z \gtrsim 20$. The monopole recovery (red) is effectively perfect, lying right on top of the input curve. For reference, we also show the SFRD of bright $M_{\rm{UV}} < -19$ galaxies (dotted black) and the \citet{Madau2014} model based on lower redshift constraints (cyan). The shaded polygons are drawn from \citet{Bouwens2023}, who provide a qualitative classification of early JWST high-$z$ galaxy candidates from ``robust'', to ``solid'', to ``possible'' from bottom to top. Note that any inference of the \textit{total} SFRD from galaxy surveys is model-dependent, in lieu of a strong detection of the turn-over. }
\label{fig:sfrd_recon}
\end{center}
\end{figure}

\begin{figure}
\begin{center}
\includegraphics[width=0.49\textwidth]{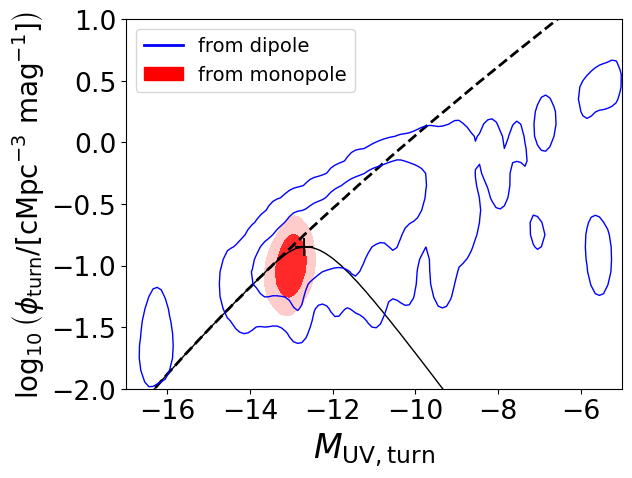}
\caption{{\bf Mock constraints on the UVLF turn-over at $z=8$.} The solid black line and cross indicates the true UVLF in model A, while the dashed black curve shows an extrapolation of the faint-end slope. A monopole measurement (red) clearly recovers the location of the turn-over to roughly one AB magnitude precision. Constraints from the dipole (blue) are suggestive of a turn-over at $-14 \lesssim M_{\rm{UV}} \lesssim -9$, but are  weaker. Though the dipole contours are broader, their angle correctly indicates a preference for a shallower UVLF than the extrapolation would suggest. }
\label{fig:uvlf_recon}
\end{center}
\end{figure}

Next, we relax the assumption of $S=1$, and fit our fiducial signal allowing the three additional parameters $M_c$, $\gamma_1$, and $\gamma_2$ to vary. Recall that these parameters allow for a possible up-turn in the faint-end of the UVLF prior to any eventual decline. Given the additional degeneracies with $M_{\rm{turn}}$, we focus on the extent to which we can recover the cosmic star formation rate density (SFRD) and turn-over in the UVLF, given by $\phi(M_{\rm{UV,turn}})$, rather than the free parameters themselves. The results of this exercise are shown in Fig. \ref{fig:sfrd_recon} and \ref{fig:uvlf_recon}.

First, in Fig. \ref{fig:sfrd_recon} we examine the recovered SFRD. Reassuringly, this more flexible model still yields a strong constraint on the SFRD. For reference we also show the parametric SFRD model from \citet{Madau2014}, based on lower redshift measurements, the fraction of the SFRD in $M_{\rm{UV}} < -19$ galaxies (dotted black), as well as three polygons indicating plausible high-$z$ star formation scenarios consistent with recent JWST results \citep{Bouwens2023}. Clearly, a measurement of the dipole would provide a powerful constraint on the \textit{total} amount of star formation at these redshifts, and so be very complementary to UVLF-based constraints, especially since UVLF measurements directly probe only relatively bright galaxies and struggle to access early phases of cosmic dawn.

Next, in Fig. \ref{fig:uvlf_recon}, we show constraints on the faint-end of the UVLF, using the same line-style and colour conventions as in Fig. \ref{fig:sfrd_recon}. The monopole (red) provides an impressive constraint on the turn-over location. The dipole constraints are weaker, as expected from the posterior in the $M_{\rm{turn}}$-$r_{\rm{turn}}$ plane (see Fig. \ref{fig:constraints_21cm_other} and associated text). However, encouragingly the angle of the posterior distribution clearly indicates a preference for a departure from a smooth power-law extrapolation below $M_{\rm{UV}} \sim -12$. As a result, a dipole measurement could provide evidence of a turn-over, without necessarily yielding a tight constraint on its location.

\begin{figure*}
\begin{center}
\includegraphics[width=0.98\textwidth]{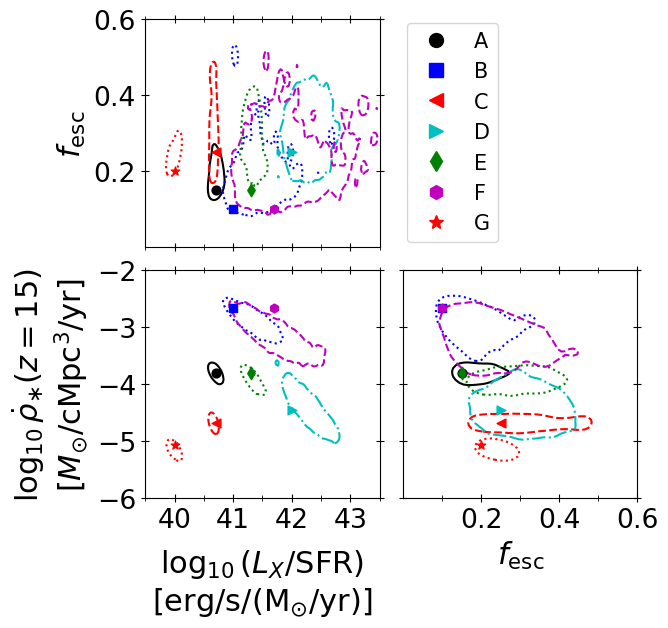}
\caption{{\bf Summary of constraints for a suite of mock 21-cm dipole signals,} focusing on the $z=15$ SFRD, $L_X/\rm{SFR}$ relation, and UV escape fraction. Contours shown are 68\% confidence regions. Constraints are weakest for scenarios with shallow dipole signals like model F (dashed magenta, hexagon), corresponding to the upper right corner of $L_X/\rm{SFR}$--SFRD space (bottom left panel). On the other hand, strong dipole signals like model G (dotted red, star), which are currently disfavoured by 21-cm power spectrum limits from HERA, are very well constrained and live in the opposite corner of parameter space. In general, the constraints are compelling and unbiased.}
\label{fig:constraints_v_rlz}
\end{center}
\end{figure*}

Finally, we broaden our exploration of the dipole's constraining power to our full model suite (introduced in Fig. \ref{fig:story} and Table \ref{tab:params}), focusing on the dipole's ability to constrain the total amount of star formation at $z=15$, the UV photon escape fraction, and the efficiency of X-ray photon production. The results are shown in Fig. \ref{fig:constraints_v_rlz}.

In general, Fig. \ref{fig:constraints_v_rlz} shows that the true input values (indicated with plot symbols) are recovered at the $1 \sigma$ level. In a few cases, the true input value is on the edge of the 68\% confidence region (models B and D) or just beyond (model F). These models are the most difficult to constrain with the dipole, as they have high SFRD and/or $L_X/\rm{SFR}$ values, which weaken the signal. However, models with $L_X/\rm{SFR} \lesssim 10^{42}$ and $\rm{SFRD}(z=15) \lesssim 10^{-3}$ are constrained extremely well. The escape fraction is more difficult to constrain, as the reionization piece of the 21-cm signal is a fairly smooth and gradual function of frequency, but in many cases these constraints would still dramatically reduce the range of viable models.

Though Fig. \ref{fig:constraints_v_rlz} is not completely exhaustive, it gives a good sense of our sensitivity to models spanning a wide range of possibilities -- 4-5 orders of magnitude in both SFRD and $L_X/\rm{SFR}$, and a factor of a few in $f_{\rm{esc}}$. The latter may not seem terribly impressive, however, any independent handle on $f_{\rm{esc}}$ would be most welcome at the moment given the apparent tensions between new JWST measurements and reionization constraints from the CMB and Ly-$\alpha$ forest. The severity of the tension depends sensitively on how $f_{\rm{esc}}$ depends on galaxy properties, with different observational constraints yielding significantly different reionization scenarios \citep[see, e.g.,][]{Munoz2024,Pahl2024}.

Finally, we note that the same general trends in Fig. \ref{fig:constraints_v_rlz} hold whether we plot the SFRD at $z=10, 15,$ or 20. Of course, the SFRD constraints generally get worse as we consider higher redshifts -- this is expected given the waning power of UVLFs at increasingly high redshifts, which can be seen in Fig. \ref{fig:sfrd_recon} -- but even at $z=20$ the dipole yields solid order-of-magnitude level constraints on the SFRD, which is difficult to imagine obtaining in any other way aside from 21-cm fluctuations or the 21-cm monopole.

\section{Discussion \& Conclusions} \label{sec:conclusions}
In this work, we explored the information content of the 21-cm dipole signal relative to the monopole. At first glance, one might expect the dipole to yield poor constraints on astrophysical parameters of interest given that to leading order it probes only the derivative of the 21-cm monopole, and so does not explicitly contain information about the amplitude of the signal. However, there are two situations in which the dipole could in principle still provide tight constraints on parameters: (i) if the derivative of the 21-cm monopole is sufficiently informative on its own to enable meaningful constraints, and (ii) if reasonable astrophysical priors can serve as a stand-in for actual measurements of the overall amplitude.

We find that indeed, the monopole still generally outperforms the dipole in terms of astrophysical constraining power. However, in many cases the dipole constraints are competitive, e.g., we found in Fig. \ref{fig:constraints_21cm_other} that constraints on $M_{\rm{turn}}$, $f_{\rm{esc}}$, and $L_X/\rm{SFR}$ -- three of the most sought after parameters of this era -- can be recovered very well with a dipole measurement. The outlier in this exercise was the parameter $r_{\rm{turn}}$, which controls the sharpness of any turn-over in the UVLF of galaxies, which can be bounded with a monopole measurement, but essentially becomes a nuisance parameter for dipole-based inference. Given this result, it was unsurprising to find that the dipole struggles to accurately constrain the position of any UVLF turn-over, as shown in Fig. \ref{fig:uvlf_recon}. However, so little star formation occurs below $M_{\rm{turn}}$ that the cosmic SFRD at $z \geq 10$ can still be constrained to a factor of $\sim 2$ in most cases. This is of course another key quantity in galaxy formation theory, and such a constraint would be more than sufficient to rule out broad classes of models.

Our parameterization choices, while surely flexible enough to accommodate a wide range of scenarios, may artificially bias our inference. For example, we have assumed that X-ray emission closely tracks star formation, when in reality there may be nuclear black holes growing in galaxies whose X-ray emission scales differently with galaxy properties. Such a population could have interesting signatures in the 21-cm background \citep[e.g.,][]{Tanaka2016}, especially if they are radio loud \citep{EwallWice2020}. In our current framework, large inferred values of the cosmic SFRD and/or $L_X$/SFR relation might provide early indications that a model based entirely on stellar X-ray sources is inadequate, but a more detailed treatment with multiple source populations is of course worth exploring. We will revisit this possibility in future work.

On a more technical level, there are several key modeling challenges that must be solved in order for the potential constraining power of the monopole and dipole to be fully realized. As mentioned in \S\ref{sec:models}, our two-zone model is approximate from the outset, having made several simplifying assumptions to avoid simulating the full 21-cm field. The assumption that the mean 21-cm background traces the product of the mean neutral fraction and contrast, $1-T_{\rm{CMB}}/T_S$, is likely accurate at the $\sim 10$\% level \citep{Schaeffer2024}, and there are plenty of generally-ignored shortcomings in models that can contribute at the $\sim 5-20$ mK level, e.g., marginalizing over cosmological parameters, accounting for systematic differences in stellar population synthesis modeling, and halo mass function uncertainties \citep{Mirocha2021,Mirocha2021b,Greig2024}. Provided these problems can be solved, the only fundamental limit is cosmic variance, which is comparable to the single-channel uncertainties assumed in this work $\sim 0.1$ mK \citep{Munoz2021CV}.

Despite these challenges of interpretation, the internal consistency check offered by the dipole is robust. We showed that one can reliably reconstruct the monopole from a dipole measurement, even if the model used in the fit is different from the true underlying model used to generate the mock data. For a measurement with $\sim 0.1-1$ mK uncertainties on the dipole, the position of the absorption feature in the monopole is recovered to a few MHz and $\sim 20-30$ mK (at 68\% confidence) for each of the three models we consider. Yet more flexible parameterizations are likely worth exploring, e.g., splines \citep{Pritchard2010,Harker2012} or the ``flex-knot'' approach \citep{Heimersheim2023,Shen2023}. Purely phenomenological approaches like these will need some additional prior, e.g., that the signal vanishes at $z \sim 6$ and $z \sim 30$, in order to constrain the overall amplitude, which the phenomenological $\tanh$ model \citep{Harker2016} sidesteps by flexibly parameterizing physical quantities rather than the signal directly.

In conclusion, the 21-cm dipole signal from $z \gtrsim 6$ offers a very interesting target for future observations. It should not be deemed valuable only as a cross-check on monopole measurements: it can in principle provide tight constraints on key parameters of galaxy formation models entirely on its own. This conclusion has been drawn from a relatively idealized forecast in order to set some initial expectations. A follow-up paper, Anderson et al., in prep., will present a schematic instrument and survey design strategy, including detailed treatments of astrophysical foregrounds and instrumental effects, in order to better assess the dipole's constraining power in a more realistic setting.

\section*{Acknowledgements}
The authors thank Mike Seiffert, Andrew Romero-Wolf, Yun-Ting Cheng, and Joe Lazio for helpful feedback on this work. J.M. was supported by an appointment to the NASA Postdoctoral Program at the Jet Propulsion Laboratory/California Institute of Technology, administered by Oak Ridge Associated Universities under contract with NASA. C.A. and T.-C.C. acknowledge support by NASA ROSES grant 21-ADAP21-0122 and the JPL 7X formulation office. Part of this work was done at Jet Propulsion Laboratory, California Institute of Technology, under a contract with the National Aeronautics and Space Administration (80NM0018D0004). The authors acknowledge the Texas Advanced Computing Center\footnote{\url{http://www.tacc.utexas.edu}} (TACC) at The University of Texas at Austin for providing computational resources that have contributed to the research results reported within this paper.
\section*{Data Availability}
The data underlying this article is available upon request.



\bibliographystyle{aasjournal}
\bibliography{main} 


\label{lastpage}
\end{document}